\documentclass[aps,twocolumn,showpacs,preprintnumbers,floatfix]{revtex4}
\usepackage{graphicx}
\usepackage{epsfig}
\usepackage{dcolumn}
\usepackage{bm}
\begin{document}

\title{Radially Excited States of 1P Charmonia and X(3872)}
\author{
Ying Chen~$^{ab}$, Chuan Liu~$^{c}$, Yubin Liu~$^{d}$, Jianping
Ma~$^{e}$, Jianbo~Zhang~$^{f}$\\
(CLQCD Collaboration)} \affiliation{ $^a$~Institute of High Energy
Physics, Chinese Academy of Sciences, Beijing
100049, P.R. China \\
$^b$~Theoretical Center for Science Facilities, Chinese Academy of Sciences, Beijing 100049,
 P.R. China\\
$^c$~School of Physics, Peking University, Beijing 100871, P.R. China\\
$^d$~School of Physics, Nankai University, Tianjin 300071, P.R. China\\
$^e$~Institute of Theoretical Physics, Chinese Academy of
Sciences, Beijing 100080, P.R. China\\
$^f$~Department of Physics, Zhejiang University, Hangzhou,
Zhejiang 310027, P.R. China }

\begin{abstract}
The excited states of charmonia are numerically investigated in quenched lattice
QCD with improved gauge and Wilson fermion actions formulated on
anisotropic lattices. Through a constrained curve fitting algorithm, the masses of the 
first excited states in $0^{++}$, $1^{++}$, and $1^{+-}$ channels are determined 
to be 3.825(88), 3.853(57), and 3.858(70) GeV, respectively. Furthormore, a node
structure is also observed in the Bethe-Salpeter amplitude of the 
$1^{++}$ first excited state. These observations indicate that
X(3872) could be the first radial excitation of $\chi_{c1}$.
\end{abstract}
 \maketitle
The narrow charmonium-like X(3872) (with width $\Gamma < 2.3$ MeV)
was first observed by Belle in the exclusive decay
$B^{\pm}\rightarrow K^{\pm}X\rightarrow K^{\pm}\pi^+\pi^-
J/\psi$~\cite{x3872-1}, and has been confirmed by
CDFII~\cite{x3872-2}, D\O~\cite{x3872-3}, and BaBar~\cite{x3872-4}
in three decay and two production channels. Even though its $J^{PC}$
quantum numbers have not been finally established, the present
experimental data strongly favor that it is a $1^{++}$
state~\cite{x3872-5} for the following reasons. First, the decay
mode $X(3872)\rightarrow \gamma J/\psi$ observed recently by Belle
requires the charge conjugation of $X(3872)$ to be
positive~\cite{x3872-6}. The possibility with $J^{PC}=0^{++}$ and
$0^{-+}$ can be ruled out based on the angular correlations in the
$\pi^+\pi^- J/\psi$ system~\cite{x3872-1} and that of $2^{-+}$ or
$1^{-+}$ is also strongly disfavored according to the dipion mass
distribution~\cite{x3872-6}. Second, the bound $\Gamma(e^+e^-)\,{\rm
Br}(X\rightarrow \pi^+\pi^- J/\psi)<10$ eV at $90\%$ C.L., obtained
with the data collected by BES at $\sqrt{s}=4.03$ GeV~\cite{bes},
implies that it is unlikely a $1^{--}$ vector state.

\par
Since the discovery of $X(3872)$, there have been many theoretical
studies. Given its quantum number $J^{PC}=1^{++}$, a natural
assignment of the state is the first radial excitation of $1P$
charmonium state $\chi_{c1}$. However, there are two main
difficulties for this interpretation. One is its tiny decay width
relative to other charmonium states, another is that it lies roughly
100 MeV lower in mass than the prediction of the quark
model~\cite{swanson}. These difficulties motivate many
non-charmonium explanations of $X(3872)$, such as
hybrids~\cite{hybrid}, glueballs~\cite{glueball}, diquark
clusters~\cite{tetra}, and molecular states~\cite{molecule}.
\par
Although the non-relativistic quark model  is successful
for heavy quark bound states, it is known that relativistic effects
can be important for charmonia , since the charm quark is not heavy
enough. In particular, these effects, which are not taken into
account in the model, can be more important for higher excited
states. It is therefore more desirable to study charmonia with a
relativistic lattice QCD formalism, which includes all the
relativistic effects. In contrast to the study of the ground states
of charmonia, such as the $1S$ and $1P$ states, their excited states
have not been investigated as much in the formalism of lattice QCD,
even though they are more interesting in the present era when many
new heavy mesons of open-charm and closed-charm are observed. The
major obstacle is that the extraction of the excited states remains
 a challenge in Monte Carlo simulations. In this work, we
investigate the relevant charmonium spectra in quenched lattice QCD
and focus on the derivation of the first (radially) excited states
through the sequential empirical Bayes method (SEB)~\cite{bayes}
advocated by $\chi$QCD collaboration, which is in the spirit of the
constrained curve fitting algorithm and has been
successfully applied to the study of nucleon excited
states~\cite{roper} and pentaquarks~\cite{penta}.
\begin{table}[h]
\caption{\label{tab:lattice} The input parameters for the
calculation. Values for the coupling $\beta$, anisotropy $\xi$,
the lattice spacing $a_s$, lattice size, and the number of
measurements are listed.}
\begin{ruledtabular}
\begin{tabular}{cccccc}
$\beta$ &  $\xi$  & $a_s$(fm) & $La_s$(fm)&
 $L^3\times T$ & $N_{conf}$ \\\hline
   2.4  & 5 & 0.222 & 3.55 &$16^3\times 80$ & 200 \\
   2.6  & 5 & 0.176 & 2.82 &$16^3\times 80$ & 200  \\
   2.8  & 5 & 0.139 & 2.22 &$16^3\times 80$ & 800  \\
\end{tabular}
\end{ruledtabular}
\end{table}
\par
We use the quenched approximation in this study. The gauge
configurations are generated by the tadpole improved gauge
action~\cite{morningstar} on anisotropic lattices with the
temporal lattice much finer than the spatial lattice, say,
$\xi=a_s/a_t\gg 1$, where $a_s$ and $a_t$ are the spatial and
temporal lattice spacing, respectively. Each configuration is
separated by 2000 heat-bath updating sweeps to avoid the
autocorrelation. The much finer lattice in the temporal direction
gives a high resolution to hadron correlation functions, such that
masses of heavy particles can be tackled on relatively coarse
lattices. The relevant input parameters are listed in Table~\ref{tab:lattice} ,
where $a_s$'s are determined from  $r_0^{-1}=410(20)$ MeV.

For fermions we use the tadpole improved clover action for
anisotropic lattices~\cite{chuan1}. The parameters in the action are
tuned carefully by requiring that the physical dispersion relations
of vector and pseudoscalar mesons are correctly reproduced
at each bare quark mass~\cite{chuan2}. The conventional interpolation
operators $\bar{\psi}\Gamma\psi$ are used for meson states with 
different gamma matrices $\Gamma$ for the specific spin-parity quantum numbers. 
After the Coulomb gauge fixing, the wall-source quark propagators are calculated under the 
anti-periodical boundary condition and are used to construct the correlation functions 
(the point-source correlation functions are found to be very noisy in 
$0^{++}$, $1^{++}$, and $1^{+-}$ channels). The bare charm quark masses at
different $\beta$ are determined by the physical mass of $J/\psi$
$m_{J/\psi}=3.097$ GeV.

\begin{figure}[htb!]
\includegraphics[height=4.0cm]{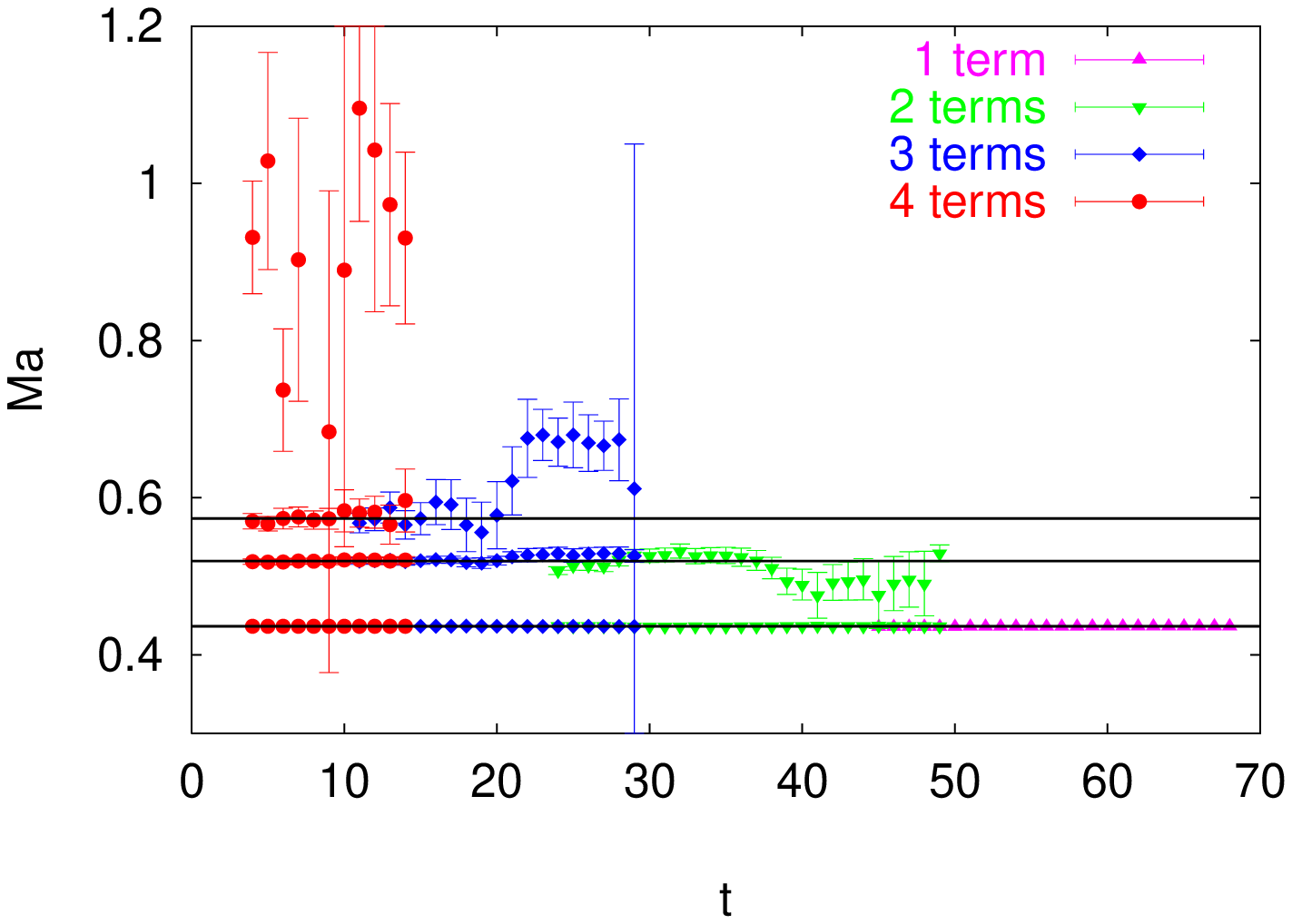}
\includegraphics[height=4.0cm]{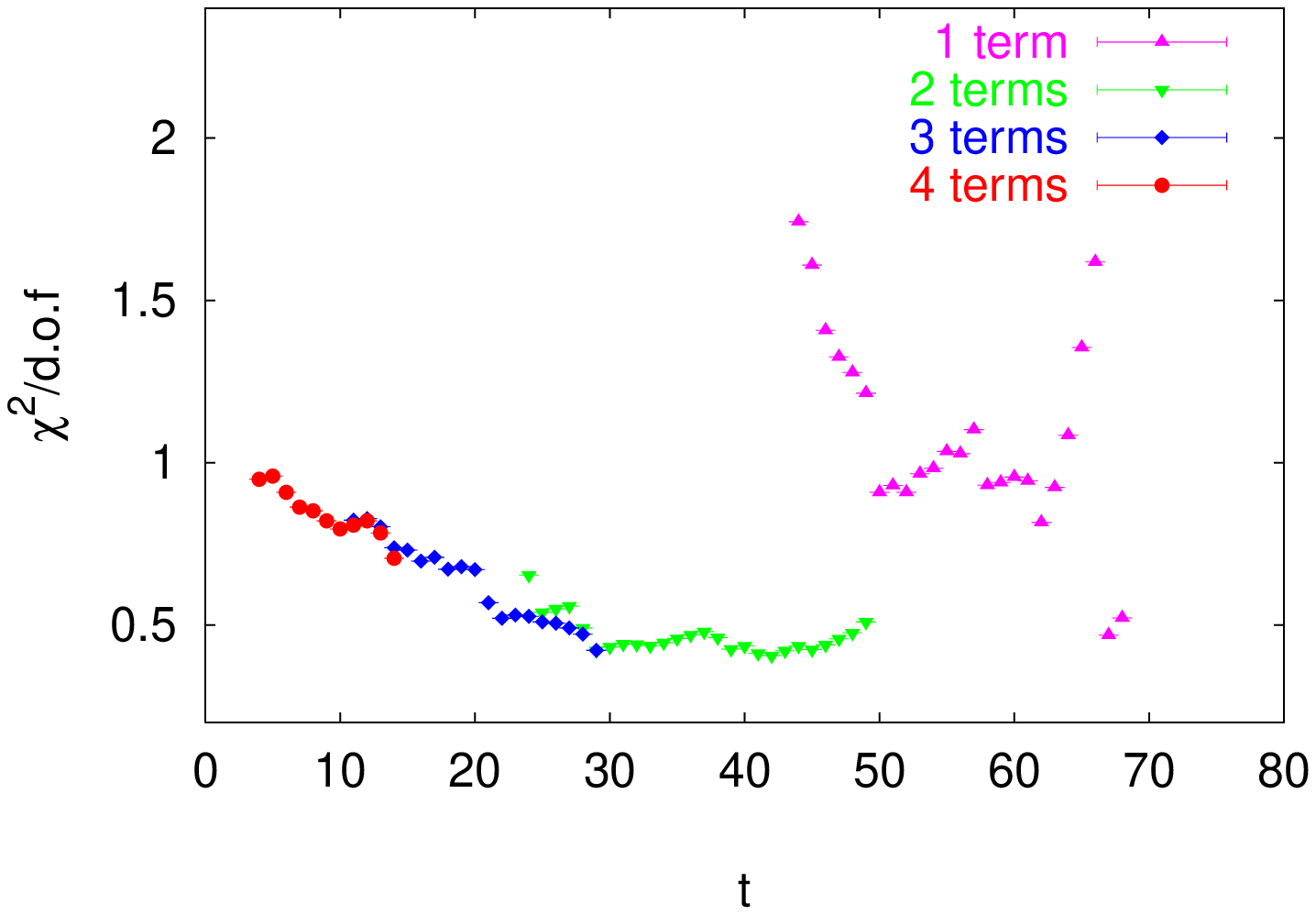}
\caption{\small The SEB fitting procedure of the vector charmonium
correlation function at $\beta=2.8$. The upper panel shows the
fitted masses using fitting models with 1-4 mass terms. The lower
panel shows the $\chi^2/d.o.f$ with the change of fitting range. }
\label{jpsi}
\end{figure}

\par
A typical correlation function takes the function form,
\begin{equation}\label{fit-model}
C(t)= \sum\limits_{i=1} W_i e^{-M_i t},
\end{equation}
where $W_i$ and $M_i$ are the spectral weight and the mass of the
$i$-th state, respectively. If one fits the correlation function
using this function form with multiple mass terms directly through
the conventional maximal likelihood method, usually one gets a
poor result due to the complicated parameter space and the result
depends crucially on the the choice of the initial values of the
parameters. A constrained curve fitting algorithm can help if one
can obtain reasonable priors of the parameters that will be
fitted~\cite{constrain}. In this work, we apply the sequential
empirical Bayes method (SEB) in the data analysis. We will outline
the main ideas of this method here and more details can be found
in Ref.\cite{bayes}.
\par
It is obvious in Eq.~(\ref{fit-model}) that the contribution of the
lowest state dominates the correlation function when $t$ is large
enough, because the relative contribution of higher states to the
ground state damps exponentially as $\propto \exp(-\Delta M_{ji}t)$,
where $\Delta M_{ji}=M_j-M_i$. Intuitively, if $W_1$ and $M_1$ of
the first state are correctly derived in a time region
$[t_1,t_{max}]$, one can treat them as priors for the ground state
and then do the two-mass-term fit in a larger time range
$[t_2,t_{max}]$ with $t_2 < t_1$. The criterion for including the
second mass term is the observation of a sharp jump of
$\chi^2/n.o.f$, which is a signal that a single exponential cannot
describe the data well, when $t_1$ is decreased further. This
procedure is repeated by adding more states  until the data points
are exhausted. This is the basic fitting procedure of SEB method.
Generally speaking, the last state can not be taken as the realistic
one because it includes almost all the contaminations from higher
states. Figure~\ref{jpsi} illustrates the SEB fitting procedure in
the vector channel. The upper panel shows the fitted masses using
1-4 mass terms in the fit function, while the lower panel shows the
$\chi^2$ per degree of freedom ($\chi^2/d.o.f$). We perform one-mass
fit in the time range $[t_1,t_{max}]=[50,68]$, and then add the
second mass term at $t=49$ where a sharp increase of $\chi^2/d.o.f$
is observed. In the two-mass fit, the fitted mass of the second
state becomes more and more stable when decreasing $t_2$, and the
$\chi^2/d.o.f$ does not change much all the way down to $t=30\equiv
t_2$. At $t=29$ $\chi^2/d.o.f$ begins to climb up and we add the
third state from there to do a three-mass fit. Similarly, we add the
fourth state from $t=14$ down to $t=4$ and find that the four-mass
fit can model the correlation function very well in the whole time
range $[t_{min},t_{max}]=[4,68]$ with a $\chi^2/d.o.f<1$. One can
see from the figure that not only the mass of the ground state keeps
constant, but also the masses of the second and even the third state
are also very stable after the fourth state is included in the
fitting model. The best fit results of the masses for the three
lowest states are 0.4367(2), 0.517(3), and 0.577(10), which
correspond to the physical masses $3099(2)$MeV, $3669(22)$MeV, and
$4084(90)$MeV, respectively, given $a_s=0.139$fm. These results,
which are in good agreement with the experimental values, illustrate
the efficacy of the SEB method.
\begin{figure}[htb!]
\includegraphics[height=4.0cm]{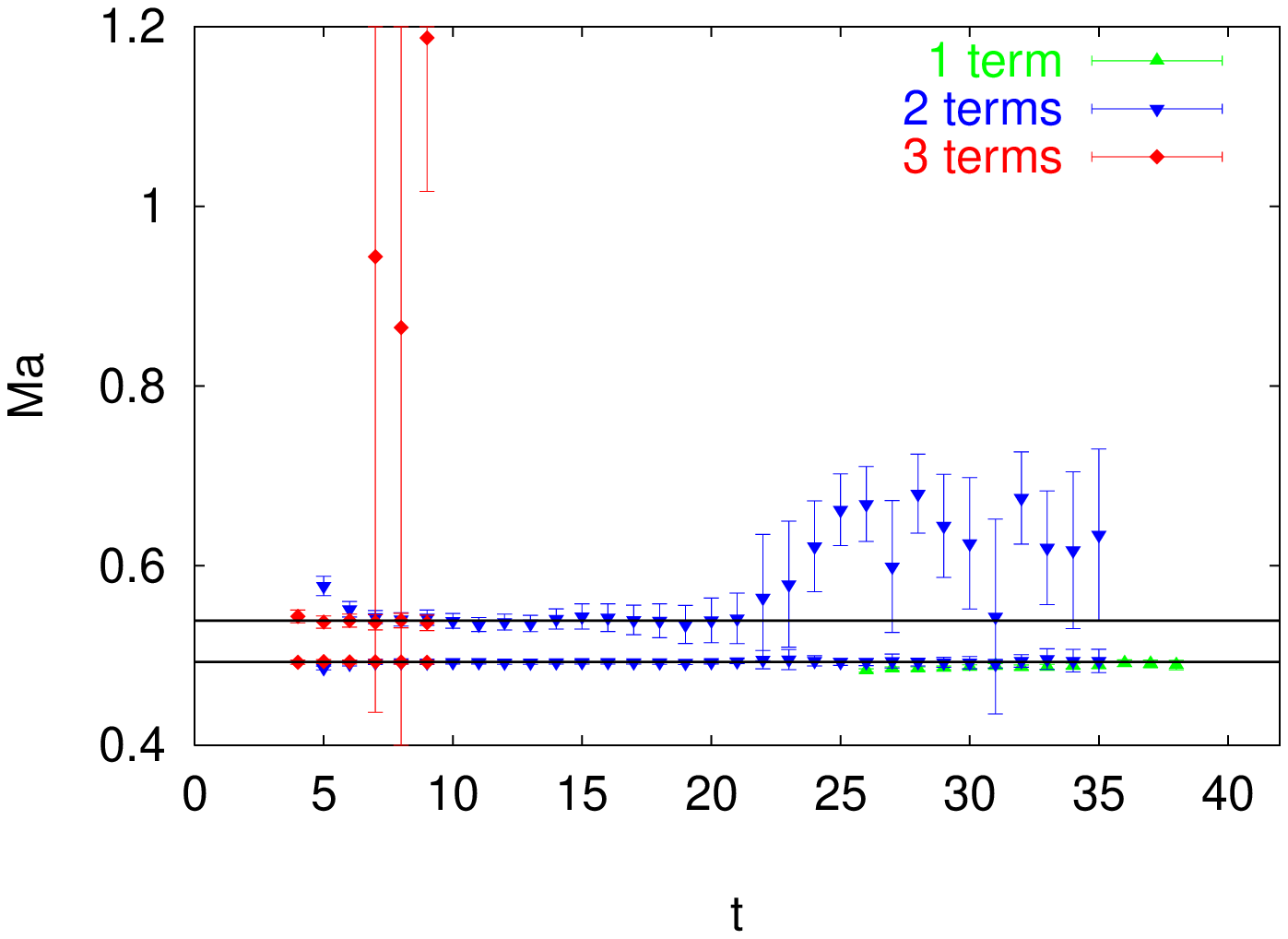}
\includegraphics[height=4.0cm]{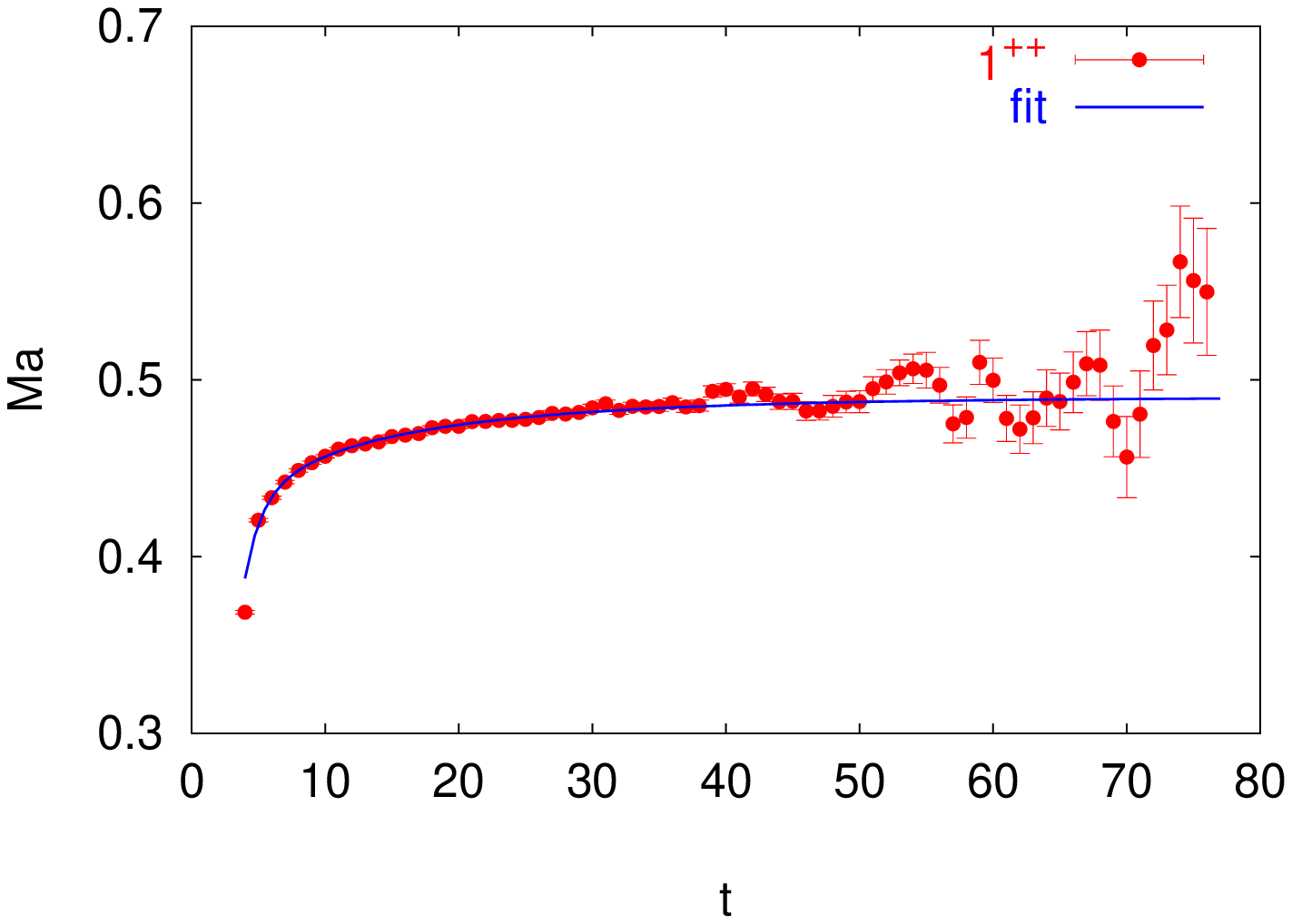}
\caption{\small The fitting procedure in $1^{++}$ channel. The
upper panel is similar to that of Figure~\ref{jpsi} but shows the
three-mass-term fit. The lower panel illustrates the effective
mass plateau $m_{\rm eff}=\ln C(t)/\ln C(t+1)$, where the data
points are the simulations results while the curves are the
fitting function plotted using the best fitted parameters.}
\label{1++}
\end{figure}
\par
Now we come to the analysis of the correlation functions in
$0^{++}$, $1^{++}$, and $1^{+-}$ channels. It is found that
these correlation functions are much noisier than those in vector (V) and
pseudoscalar (PS) channels, such that we can only carry out maximally 
three-mass-term fits to them in the practical data analysis. 
Figure~\ref{1++} shows the fitting procedure 
for the $1^{++}$
charmonium at $\beta=2.8$. The upper panel is similar to that of
Figure~\ref{jpsi}, but shows the three-mass-term fit, where one
can see that the second state is very stable after the inclusion
of the third state in the fitting model. The lower panel
illustrates the effective mass plateau $m_{\rm eff}=\ln C(t)/\ln
C(t+1)$, where the data points are the simulation results while
the curve is the three-mass fitting function with the best-fit
parameters. The good agreement of the curve and the data points
manifests the quality of the three-mass fit. The fitting procedure
for the $0^{++}$ and $1^{+-}$ channels is similar.
\par
We also carry out the same calculations at $\beta = 2.4$, $2.6$ with
smaller statistics (200 configurations each) to consider the
continuum extrapolation. Due to the larger lattice spacings at these
two $\beta$'s as shown in Table~\ref{tab:lattice}, the largest time
for the fit windows in parity-positive channels is roughly $t_{\rm
max}=20$ and $30$ respectively. For both $\beta$, the correlation
functions can be well fitted by only two mass terms in the full time
windows and the third mass terms are found to be marginal. This
implies that the fitted first excited states may have significant
contaminations from higher states. The continuum limits are obtained
by the linear extrapolation in terms of $a_s^2$.
Table~\ref{threebeta} lists the best-fit masses of the ground and
the first excited states at the three $\beta$'s and their continuum
extrapolations. The extrapolation values of the $P$-wave ground
states are in agreement with previous works~\cite{chen, cp-pacs} but
lower than the experiment values.
The results of the extrapolated masses of the first excited
states should be taken with caution because
the masses at the two small $\beta$'s can have large systematic errors as mentioned above.
\par
The masses of the excited masses obtained at
$\beta=2.8$ are relatively more reliable and can be treated as
approximations of their continuum value with the awareness that
the uncertainties from the finite lattice spacing are not properly
tackled. Comparing with the predictions by the non-relativistic
quark model ~\cite{swanson} (the last colume of
Table~\ref{threebeta}), the mass of the $1^{++}$ first excited
state at $\beta=2.8$ is lower than quark model expectation but
consistent with the experimental value of $X(3872)$. Even though
our result is not decisive to the assignment that $X(3872)$ be the
first radial excitation of $\chi_{c1}$ due to the large
statistical error, it indicates that the possibility that $X(3872)$ is
a conventional charmonium state can not be simply ruled out.
\begin{table}
\caption{\label{threebeta} Best-fit masses of the ground and the
first excited states of parity-positive charmonia at different
$\beta$. All errors are statistical. The errors of the continuum
limit values are from the linear extrapolation in terms of
$a_s^2$.}
\begin{ruledtabular}
\begin{tabular}{ccccc|c}
$\beta$  &     2.4     &   2.6    &    2.8   & cont. & BGS\cite{swanson}\\
\hline
$0^{++}(1P)$  & 3.466(19) &3.437(21)  &{\bf 3.410(18)}&3.376(6) & 3.424  \\
$1^{++}(1P)$  & 3.520(19) &3.507(17)  &{\bf 3.477(14)}&3.453(6) & 3.505  \\
$1^{+-}(1P)$  & 3.515(17) &3.505(17)  &{\bf 3.488(13)}&3.472(8) & 3.516  \\
\hline
$0^{++}(2P)$  & 3.816(57) &3.865(90)  &{\bf 3.825(88)}&3.857(55) & 3.852 \\
$1^{++}(2P)$  & 3.937(56) &3.887(54)  &{\bf 3.853(57)}&3.800(2) & 3.925 \\
$1^{+-}(2P)$  & 3.955(53) &3.964(71)  &{\bf 3.858(70)}&3.835(88)
 &
3.934
\end{tabular}
\end{ruledtabular}
\end{table}
\par
In order to see if the extracted masses are really those of the first excited states,
we also investigate the Bethe-Salpeter amplitudes of
charmonium states at $\beta=2.8$. In the Coulomb gauge, we split the sink
operator into two parts, with each quark field residing on different spatial sites, namely,
$O_\Gamma(x,y)=\bar{\psi}(x)\Gamma\psi(y)$,
where $\Gamma$ represents the various gamma matrices corresponding
to specific $J^{PC}$ quantum numbers and $x=({\bf x},t)$ and
$y=({\bf y},t)$. In the practical calculation, all the two-point
functions with $|{\bf x}-{\bf y}|\leq 6\sqrt{3}a_s$ are
calculated. It is found that the two-point functions exhibit a
spherical symmetry with respect to the relative displacement
between the quark and anti-quark fields, ${\bf r}={\bf x}-{\bf
y}$. Therefore the two-point functions with the same spatial
separation $|{\bf r}|$ are averaged to increase the statistics.
\par
The two-point functions calculated with the Coulomb wall sources
are actually $C_\Gamma({\bf r},t)=\sum\limits_{\bf
x}\langle 0|O_{\Gamma}({\bf x},t;{\bf
x+r},t))O_{\Gamma,W}^\dagger(0)|0\rangle$, where
$O_{\Gamma,W}(0)=\sum\limits_{\bf y,z}\bar{\psi}({\bf
y},0)\Gamma\psi({\bf z},0)$. After the integration over ${\bf x}$, we get
\begin{eqnarray}\label{wave}
C_\Gamma({\bf r},t)&=&\sum\limits_n \frac{1}{2M_n}\langle
0|O_\Gamma({\bf 0},0;{\bf r},0)|n\rangle\times \nonumber\\
&&\sum\limits_{\bf y,z}\langle n|O_\Gamma^\dagger ({\bf y},0;{\bf
z},0)|0\rangle {\rm e}^{-M_n t}\nonumber \\
&=&\sum\limits_{n} \Phi_n(r){\rm e}^{-M_n t}~~(n=1,2,\ldots),
\end{eqnarray}
where $M_n$ is the mass of the $n$-th state and $\Phi_n(r)$ the BS
amplitude (up to an irrelevant prefactor) we would like to extract.
\begin{figure}[htb!]
\includegraphics[height=5.0cm]{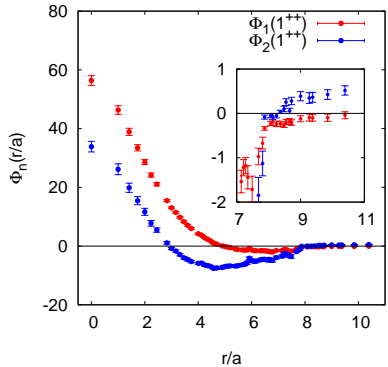}
\caption{$\Phi_n(r)$'s for the ground ($n=1$) and the
excited state ($n=2$) in the $1^{++}$ channel. The mini-panel
shows the enlarged plot around $r/a =8$.} \label{wvf}
\end{figure}
\par
The $r$-dependent BS amplitudes $\Phi_n(r)$  have direct connections with
the non-relativistic wave functions of heavy quarkonia~\cite{mocsy,bodwin}.
Specifically, for the $P$-wave charmonia,  $\Phi_n(r)$ is related approximately
to the radial wave function $R_n(r)$ as
\begin{equation}
\Phi_n(r)\sim C R_n'(r),
\label{connection}
\end{equation}
where $R_n'(r)$  is the $r$-derivative of $R_n(r)$ and $C$ a factor
irrelevant to the discussion here.  In this work, $\Phi_1(r)$ and $\Phi_2(r)$ are 
obtained from a constrained-curve-fitting algorithm by assuming the fitted
masses of the ground and the first excited states at $r=0$ as
known parameters, say, priors. Plotted in Figure~\ref{wvf} are
$\Phi_n(r)$'s of the ground ($n=1$) and the first excited state ($n=2$) in the 
$1^{++}$ channel. It is obviously seen in the figure that  $\Phi_1(r)$ has one
node at $r/a \sim 4.8 $ and $\Phi_2(r)$ has two nodes
at $r/a\sim 2.5$ and $ r/a \sim 8$. These behaviors are qualitatively
in agreement with the theoretical anticipation described by Eqn.(\ref{connection})
if both states correspond to the $1P$ and $2P$ states of $1^{++}$ charmonium:
the unique node of  $\Phi_1(r)$ corresponds to the unique maximum of $1P$
radial wave function, the two nodes of $\Phi_2(r)$ manifest the
two extrema,  and necessarily a radial node in-between,  of the $2P$ radial wave function.
\par
To summarize, we have carried out a quenched lattice study of the
excited states of charmonia on anisotropic lattices. In our
study, a relativistic formalism of lattice QCD is employed and
hence the relativistic effects are included. Using SEB, which is
in the spirit of the constrained curve fitting method, we can
derive reliably the masses of the ground states and the first
excited states of charmonia in the $0^{++}$, $1^{++}$, and
$1^{+-}$ channels. We obtain a mass of $3853(57)~\,{\rm MeV}$ for
the first excited state of the $1^{++}$ charmonium, which is lower
than the conventional quark model predictions but consistent with
the measured mass of $X(3872)$ within statistical errors
(systematic errors from the quenched approximation have not been
considered yet). We have also observed a radial node of the
wave function of the $1^{++}$ first excited state. All these
results support the possibility that $X(3872)$ be the first
radially excited state of $\chi_{c1}$.
\par  We are grateful to Prof. S.L Zhu of Peking University for
his valuable suggestions. This work is supported in part by NSFC
(Grant No.10235040, 10347110, 10575107, and 10421003) and CAS
(Grant No. KJCX3-SYW-N2). The numerical calculations were
performed on DeepComp 6800 supercomputer of the Supercomputing
Center of Chinese Academy of Sciences, Dawning 4000A supercomputer
of Shanghai Supercomputing Center, and NKstar2 Supercomputer of
Nankai University.


\begin{thebibliography}{9}
\bibitem{x3872-1} S.K. Choi {\it et al.} [Belle Collaboration],
Phys. Rev. Lett. {\bf 91}, 262001 (2003).
\bibitem{x3872-2} D. Acosta {\it et al.} [CDF II Collaboration],
Phys. Rev. Lett. {\bf 93}, 072001 (2004).
\bibitem{x3872-3} V.M. Abazov {\it et al.} [D0 Collaboration],
Phys. Rev. Lett. {\bf 93}, 162002 (2004) .
\bibitem{x3872-4} B. Aubert {\it et al.} [BABAR Collaboration], Phys. Rev. D
{\bf 71}, 071103 (2005).
\bibitem{x3872-5} K. Abe {\it et al.}, hep-ex/0505038.
\bibitem{x3872-6} K. Abe {\it et al.}, hep-ex/0505037
\bibitem{bes} C.Z. Yuan, X.H. Mo, and P. Wang, Phys. Lett. B {\bf
579}, 74 (2004).
\bibitem{swanson} See, for example, E.S. Swanson, Phys. Rept. {\bf 429}, 243 (2006).
\bibitem{hybrid} B.A. Li, Phys. Lett. B {\bf 605}, 306 (2005).
\bibitem{glueball}K.K. Seth, Phys. Lett. B {\bf 612}, 1 (2005).
\bibitem{tetra} L. Maiani, F. Piccinini, A.D. Polosa, and V.
Riquer, Phys. Rev. D {\bf 71}, 014028 (2005). T.W. Chiu and T.H.
Hsieh [TWQCD Collaboration], arXiv:hep-ph/0603207.
\bibitem{molecule} N.A. T\"{o}rnqvist, arXiv:hep=ph/0308277, F.E.
Close, and P.R. Page, Phys. Lett. B {\bf 578}, 119 (2004). S.
Pakvasa and M. Suzuki, Phys. Lett. B {\bf 579}, 67 (2004). E.S.
Swanson, Phys. Lett. B {\bf 588}, 189 (2004).
\bibitem{bayes} Y. Chen, S.J. Dong, T. Draper, I. Horv{\'a}th, F.X. Lee,
K.F. Liu, N. Mathur, C. Srinivasan, S. Tamhankar, J.B. Zhang,
hep-lat/0405001.
\bibitem{roper} N. Mathur, Y. Chen, S.J. Dong, T. Draper, I.
Horv{\'a}th, F.X. Lee, K.F. Liu, and J.B. Zhang, Phys. Lett. B
{\bf 605}, 137 (2005), hep-ph/0306199.
\bibitem{penta} N. Mathur {\it et al}., Phys. Rev. D {\bf 70},
074508 (2004).
\bibitem{morningstar} C.J. Morningstar and M. Peardon, Phys. Rev. D
{\bf 56}, 4043 (1997).
\bibitem{chuan1}Chuan Liu, Junhuan Zhang, Ying Chen, J.P. Ma,
Nucl. Phys. B {\bf 624}, 360 (2002).
\bibitem{chuan2} S. Su, L. Liu, X. Li, and C. Liu, Int. J. Mod.
Phys. A {\bf 21}, 1015 (2006), Chin. Phys. Lett. {\bf 22}, 2198
(2005).
\bibitem{constrain} G.P Lepage {\it et al.}, Nucl. Phys. Proc.
Suppl. {\bf 106} 12 (2002), hep-lat/0110175; C. Morningstar, Nucl.
Phys. Proc. Suppl. {\bf 109}, 185 (2002), hep-lat/0112023.
\bibitem{cp-pacs} M. Okamoto {\it et al.} [CP-PACS], Phys. Rev. D
{\bf 65}, 094508 (2002).
\bibitem{chen} P. Chen, Phys. Rev. D {\bf 64}, 034509 (2001).
\bibitem{mocsy} \'{A}. M\'{o}csy and P. Petreczky, Phys. Rev. D
{\bf 73}, 074007 (2006).
\bibitem{bodwin}G.T. Bodwin, E. Braaten, and G.P. Lepage, Phys. Rev. D
{\bf 51}, 1125 (1995); {\bf 55}, 5853(E) (1997).

\end{thebibliography}
\end{document}